**Tailoring 2D Fast Iterative Filtering algorithm for low-contrast optical fringe pattern preprocessing**

Mikołaj Rogalski,[1*] Mateusz Pielach,[2] Antonio Cicone,[3,4,5] Piotr Zdańkowski,[1] Luiza Stanaszek,[6] Katarzyna Drela,[7] Krzysztof Patorski,[1] Barbara Lukomska,[6] Maciej Trusiak,[1,8]

[1]Warsaw University of Technology, Institute of Micromechanics and Photonics, Warsaw, Poland

[2]Institute of Physical Chemistry, Polish Academy of Sciences, Kasprzaka 44/52, 01-224 Warsaw, Poland

[3]DISIM, Università degli Studi dell'Aquila, Via Vetoio n. 1, 67100 L'Aquila, Italy

[4]Istituto di Astrofisica e Planetologia Spaziali, INAF, Roma, Italy

[5]Istituto Nazionale di Geofisica e Vulcanologia, Roma, Italy

[6]NeuroRepair Department, Mossakowski Medical Research Institute, Polish Academy of Sciences, Warsaw, Poland

[7]Medical Research Agency, Stanislawa Moniuszki 1a St., 00-014 Warsaw, Poland

[8]e-mail: maciej.trusiak@pw.edu.pl

*Corresponding author: mikolaj.rogalski.dokt@pw.edu.pl

**Retrieving object phase from the optical fringe pattern is a critical task in quantitative phase imaging and often requires appropriate image preprocessing (background and noise minimization), especially when retrieving phase from the single-shot fringe pattern image. In this article, for the first time, we propose to adapt the 2D Fast Iterative Filtering (FIF) method for fringe pattern decomposition and develop a novel version of FIF called the 2D fringe pattern Fast Iterative Filtering (fpFIF2), that is tailored for fringe pattern preprocessing. We show the positive influence of fpFIF2 onto fringe pattern filtering comparing to the previous 2D FIF implementation regarding processing speed, quality, and usage comfortability. We also compare the fpFIF2 with other state-of-the-art fringe pattern filtering methods in terms of aiding the Hilbert spiral transform method in phase retrieval. Employing numerical simulations and experimental fringe analysis, we prove that fpFIF2 outperforms reference methods, especially in terms of low-fringe-contrast phase reconstruction quality and decomposition time.**

## 1. Introduction

Optical metrology is a field of science that, with the use of light, enables precise and non-invasive sample measurement [1,2]. In many of its techniques (e.g., interferometry, holography, moiré, structured illumination), sample information is retrieved from the recorded coherently/incoherently generated fringe pattern. Usually, from this pattern, the object phase map is computed containing the information about how much the light is delayed when passing through or reflecting/scattering from the sample. The phase may then be directly converted into the object thickness (geometry) or refractive index distribution. Quantitative phase imaging (QPI) [3,4], as one of the pivotal newly emerged optical measurement techniques, is used for microscopic imaging of transparent objects enabling, e.g., label-free cells examination crucial in many biomedical applications.

Up to this day, many methods have been proposed for phase retrieval from optical fringe patterns [5,6], which may be divided into two groups: single-frame and multi-frame methods. The multi-frame techniques are considered the most robust ones (e.g., temporal phase shift [5] or principal component analysis methods (PCA) [7]). They require at least 3 phase-shifted fringe patterns, which are usually obtained by introducing some precise mechanical movement inside the interferometric setup (e.g., with the use of piezo-mechanical actuators). This obviously increases the system's cost, complexity, and size, making it environmentally sensitive and cumbersome to implement in non-laboratory conditions. Moreover, because of the time delay between collecting each frame, multi-frame methods are not suitable for measuring dynamically changing samples. Those disadvantages are avoided in single-frame methods (e.g., Fourier transform (FT) [6,8] or spatial carrier phase shift techniques [5,9,10]). However, they require high-frequency fringe patterns, which generally limit the lateral resolution of the recovered phase image. A different approach to single-frame phase decomposition is the Hilbert spiral transform method (HST) [11–14], which synthetically generates a second, π/2-phase-shifted fringe pattern and then uses both: collected and synthetic image to straightforwardly retrieve sought phase information upon analytic signal demodulation. The HST does not require high fringe pattern carrier frequency and, therefore, may provide a higher phase space-time-bandwidth product than other single-frame methods [15]. However, the HST method requires the fringe pattern image to be noise- and background-free to work correctly. Therefore, the HST algorithm is often preceded by some numerical denoising and detrending techniques. Their role is pivotal, especially in low signal-to-noise ratio (SNR) regimes, often encountered when low contrast fringes are generated as a result of coherence/intensity/polarization mismatch of interfering beams. Hence, low SNR considerably limits the QPI accuracy. To increase the QPI applicability, we will numerically tackle the important problem of preprocessing low SNR fringe patterns.

Currently, there are two popular solutions for adaptive fringe pattern preprocessing to be considered. First of them is the bidimensional empirical mode decomposition (BEMD) [16–21] algorithm, which can decompose the input image into several bidimensional intrinsic mode functions (BIMFs), each one encompassing different spatial frequency components contributing to the image of interest on different intrinsic scales. Because of that, it requires an additional algorithm [19,20] or human interaction to select which of those components contain sought fringe information. Moreover, the BEMD exhibits mode splitting (single component spreads onto several consecutive modes) and mode mixing (two or more components are present in a single mode) and thus does not separate the frequency components between modes well, resulting in noisy fringes and fringe pattern information leakage to the noise or background-dominated modes. Therefore, the BEMD is often aided by auxiliary denoising algorithms like state-of-the-art block matching and 3D filtering (BM3D) [22–24]. The second well-established solution is the variational image decomposition (VID) [15,25–27] method. Contrary to the BEMD, VID always returns only three image components (background, fringes, and noise terms), making it straightforward to automatize. It also can separate the image components with less information leakage than in the BEMD method. However, the VID method also has drawbacks. Its performance is dependent on several input parameters and the decomposition time may be slower even by two orders of magnitude than in the BEMD method [27].

Both methods described above have thus disadvantages, which reduce their usage area and still constitute a gap in state-of-the-art which is required to be filled. Recently, an attractive alternative signal decomposition method called Fast Iterative Filtering (FIF) emerged [28–32]. The FIF principle of operation is similar to the BEMD one. Both techniques use the moving average to decompose the signal into several different-frequency modes. The critical difference is in the way this average is computed. In BEMD, it is done by generating image envelopes using spline interpolation [33,34] or statistical approximation techniques [35]. These approaches do not guarantee mathematical convergence a priori and make the algorithm unstable in the presence of noisy signals [36]. Whereas in FIF, the moving average is calculated as a convolution between signal and a tailored filter, which makes FIF mathematically convergent [32]. Furthermore, according to 1D comparisons, FIF is more stable (provides more physically meaningful components in a predictable and fully controllable way) and less noise-sensitive than empirical mode decomposition [37].

The FIF algorithm has already been implemented in 2D and called the Multidimensional Fast Iterative Filtering (FIF2) [38–40]. However, this algorithm was developed as a general 2D signal decomposition method. Hence, it does not consider the fringe pattern properties such as its spatial periodicity, the frequency variation across the image or the fringes continuity, and because of that, it is not well suited for fringe pattern decomposition. When trying to use FIF2 for fringe pattern decomposition, one may encounter several issues which cause the algorithm to return a very large (usually from 35 to even 55) number of BIMFs (which makes it time-consuming and hard to automatize). Moreover, the result of FIF2 decomposition is dependent on the fringe direction and two input variables ($\alpha$ and $\xi$ – see Section 2.1 for details), which have to be set by the user and usually require adjusting to mitigate mode-mixing. After insightful analysis, we diagnosed that these problems were mainly due to the non-ideal method of calculating the mask size (used to compute the signal moving average) and insufficient algorithm's outer loop stopping criterion. To reduce these issues, we have modified the FIF2 algorithm. As a result, we developed the fringe pattern 2D Fast Iterative Filtering (fpFIF2) algorithm, which is freely available online [41]. Our main improvements introduced in the fpFIF2 are:

1. applying a new local method for mask size calculation allowing fpFIF2 to perform independently of the fringe direction and removing the necessity of adjusting the input variables (Section 2.1),

2. applying additional BIMFs filtering that reduces the information leaking between modes (Section 2.2),

3. adding alternate algorithm's outer loop stopping criteria (Section 2.3).

All these modifications made the fpFIF2 algorithm generate around 3 times smaller number of modes. Furthermore, the fpFIF2 became more robust and about 25 times faster than the original FIF2 algorithm. This work presents the fpFIF2 algorithm principle of operation (Section 2) and compares it to the original FIF2 method (Section 3.1). We also compare the fpFIF2 performance with state-of-the-art enhanced empirical mode decomposition (EFEMD) [20], EFEMD assisted with BM3D (EFEMD+BM3D) [24] and unsupervised variational image decomposition (uVID) [27] methods in terms of assisting the HST algorithm in phase retrieval both on simulated (Section 3.2) and experimental (section 4) datasets. We show, that independently of the image conditions, the fpFIF2 performance is at least as good as compared algorithms, and in particular critical conditions of low fringe pattern SNR and low phase dynamic ranges (PDR), it successfully outperforms all other algorithms.

## 2. Description of the proposed algorithm

The diagram describing the principle of operation of the fpFIF2 algorithm is shown in Fig. 1. Blue, violet, and yellow cells on that diagram represent our modifications that are detailed in Sections 2.1-2.3. Each iteration ($i$) of the fpFIF2's outer loop begins with calculation of the filter size ($m_i$) from the currently processed image ($I_i$; in the first iteration, the $I_i$ is the input image) – see Section 2.1. Next, two stopping criteria of the algorithm's outer loop (see Section 2.3) are checked and if both are not fulfilled, the image Fokker-Planck filter ($w_i$) is generated based on $m_i$ [38]. Additionally, if there is a large difference between mask sizes calculated in this and the previous iteration (if $m_i > 2 \cdot m_{i-1}$), then the auxiliary BIMF is created (blue cells in Fig. 1, see Section 2.2). Then, the $w_i$ is adjusted to the signal in the algorithm's inner loop according to Eq. (1):

$$w_i' = F^{-1}(1 - (1 - F(w_i))^q), \qquad (1)$$

where $q$ is the inner loop iteration number, $F$ and $F^{-1}$ stand for Fourier transform and inverse Fourier transform operations, respectively. The inner loop stops, when its only stopping criterion is fulfilled:

$$Stop\ if: \frac{\sum |F(I_i)|^2 \cdot |F(w_i)|^2 \cdot (1 - |F(w_i)|^2)^q}{\sum |F(I_i)|^2 \cdot (1 - |F(w_i)|^2)^q} < \delta, \qquad (2)$$

where $\sum(x)$ represents the sum of all pixels of $x$ image and $\delta$ is the user defined parameter, by default 0.001. After that, the *n-th* BIMF is created by subtracting from the $I_i$ its moving average (which is created as convolution between $I_i$ and $w_i$):

$$BIMF_n = I_i - I_i * w_i'. \tag{3}$$

At the end of the iteration, the image processed in the next iteration is created as:

$$I_{i+1} = I_i - BIMF_n \tag{4}$$

and the last outer loop stopping criterion is checked (see Section 2.3). If any of the algorithm's stopping criteria is fulfilled, the current image is set as the last BIMF.

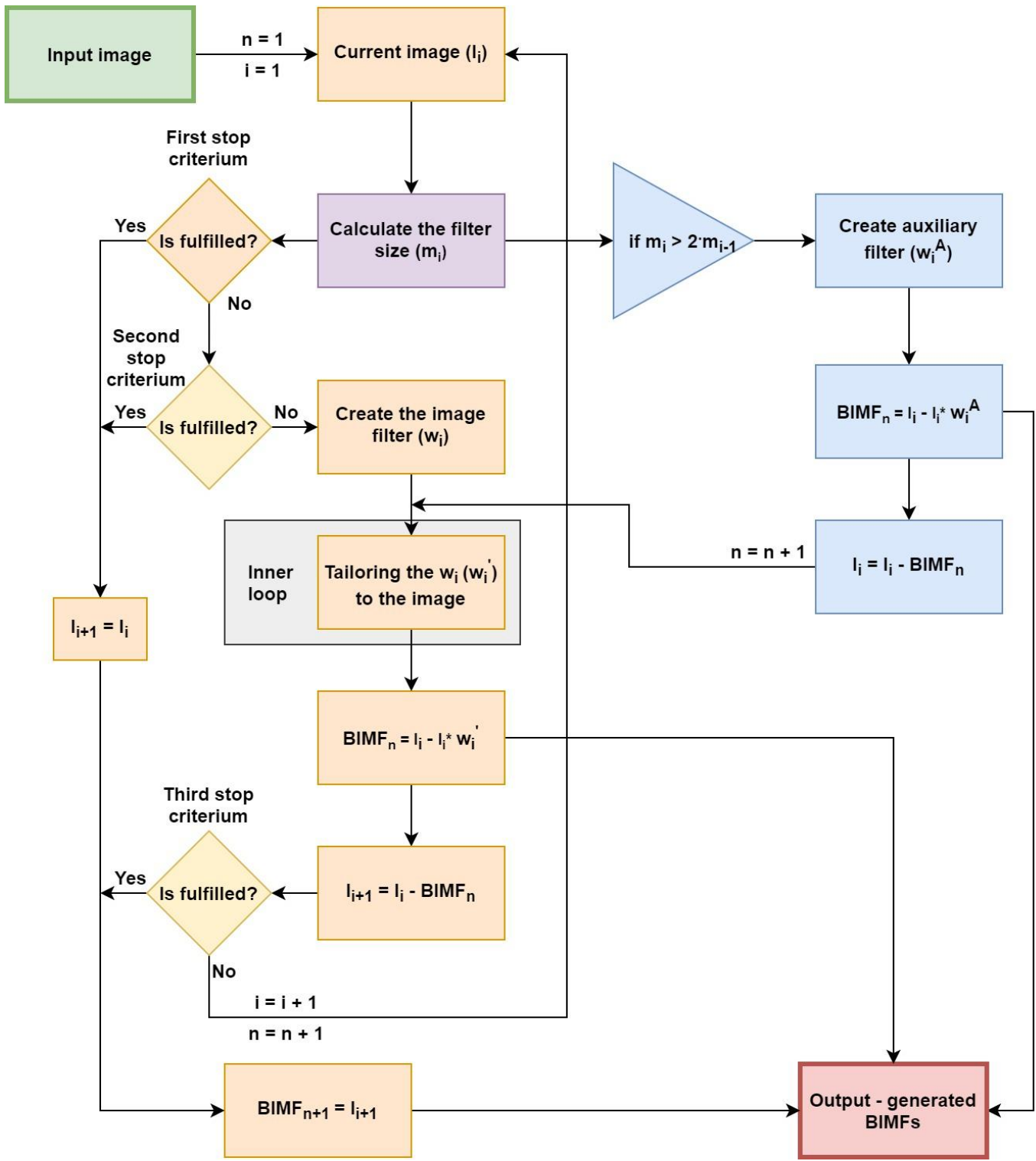

**Fig. 1.** Diagram of the fpFIF2 principle of operation. Individual modes are sequentially extracted from the input image in each iteration until one of the stopping criteria finishes the algorithm; i stands for iteration number, n for BIMF number and * represents the convolution operation.

*2.1. Filter size calculation and boosting the computation time*

In the FIF2 algorithm, the mask size is calculated basing on the distance between extrema, according to Eq. (5):

$$m_i = 2 \cdot \xi \cdot \varepsilon_i, \tag{5}$$

where $\xi$ is a user-defined parameter and $\varepsilon_i$ is the average distance between signal extrema, calculated as $\alpha$ percentile from the vector containing distances between extrema ($V_i$):

$$\varepsilon_i = \text{percentile}(V_i, \alpha). \tag{6}$$

However, in the FIF2, calculation of the $V_i$ is performed on the image reduced to 1D vector, which is not suitable for 2D fringe signal analysis and makes the algorithm behave unpredictably. Moreover, because of the unideal calculation of $V_i$, FIF2 performance strongly depends on $\xi$ and $\alpha$ parameters, which must be adjusted for each separate dataset to compensate the $V_i$ errors (coming from 1D vector representation of the analyzed 2D image) and therefore obtain better fringes filtration (smaller mode mixing effect).

In the fpFIF2, we kept the method of calculating $m_i$ based on Eq. (5) and Eq. (6) but we changed the method of calculating the $V_i$ into a more 2D suitable approach, see Fig. 2. In our algorithm, we firstly perform high pass Gaussian filtering on the image, $I_i$ – Fig. 2(a), to remove the image lowest-frequency components, which improves the sensitivity and accuracy of the $V_i$ array calculation. Then we calculate the extrema map, $E_i$ – Fig. 2(b), with the use of grayscale dilation. After that, we compute the Euclidean distance transform of the $E_i$ to obtain the map of distances from the extrema, $D_i$ – Fig. 2(c). Afterwards, we are looking for maxima of the $D_i$ termed as $M_i$ – Fig. 2(d) and compute the map of distances between extrema, namely $DE_i$ – Fig. 2(e), as:

$$DE_i = 2 \cdot D_i \cdot M_i. \tag{7}$$

One can understand $DE_i$ as a map containing information about the distance between neighboring extrema points in the pixels that are in the middle of the distance between those extrema, and zero values in all other pixels. Then, we could simply set non-zero values of $DE_i$ as $V_i$. However, in the case when the $I_i$ consists of several spatially separated fringe local frequency components like in Fig. 2(a), this solution would then favorize the presence of distances between higher frequencies in $V_i$. This is because of the fact that when occupying areas of the same size, the higher frequencies produce more signal extrema than lower-frequency components. To make the algorithm rely on the size of areas occupied by different fringe frequencies instead of their number of extrema, we are additionally dividing the $DE_i$ into 100 equally-sized regions. For each of these regions, we are finding the most frequently occurring non-zero $DE_i$ value to obtain a pseudo-local signal density map, which values we set as $V_i$ (Fig. 2(f)).

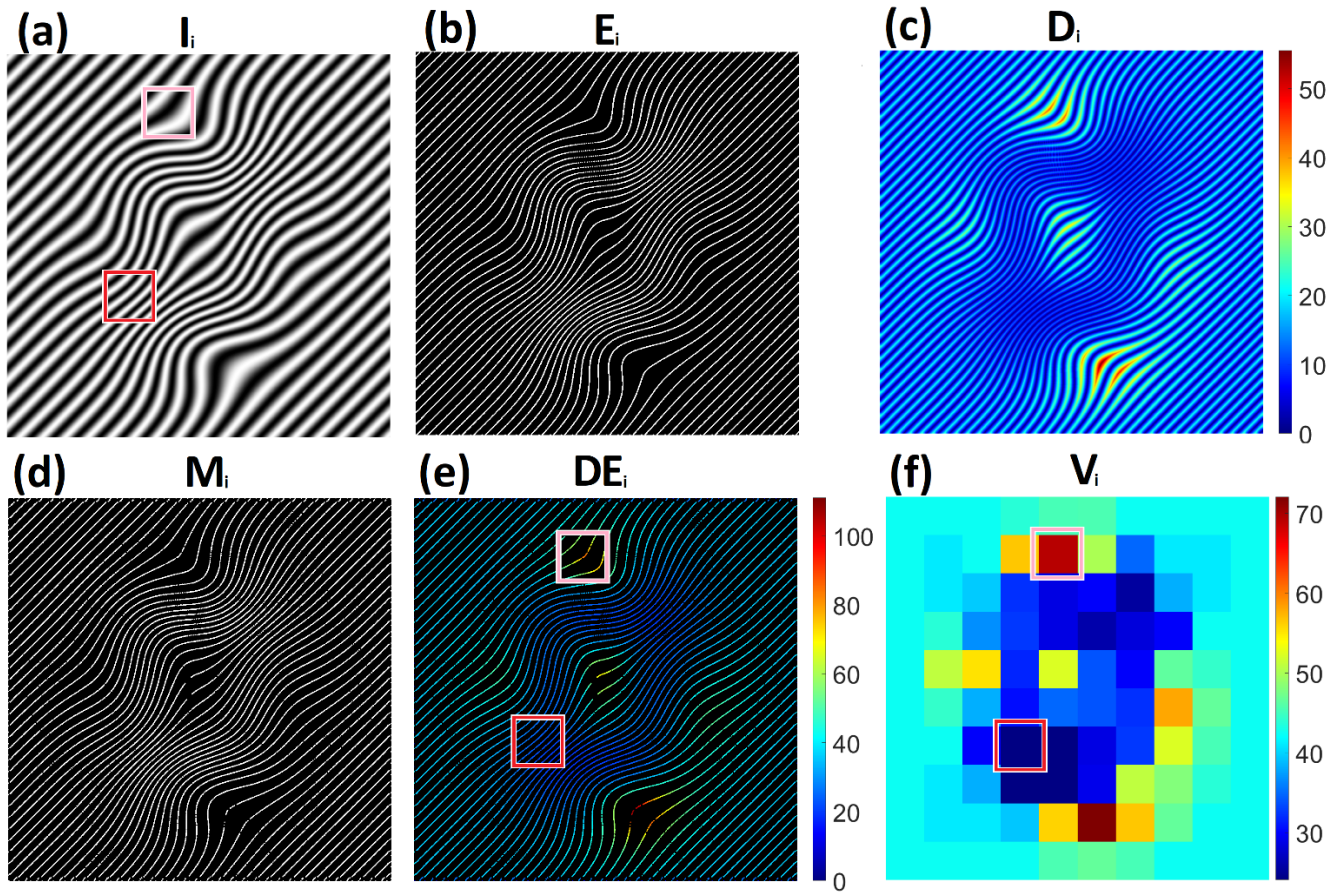

**Fig. 2.** The next steps of calculating the $V_i$ vector for an exemplary fringe pattern. (a) – input image ($I_i$); (b) – extrema map of the $I_i$ ($E_i$); (c) – map of distances from extrema ($D_i$); (d) – maxima of the distances map ($M_i$); (e) – map of distances between extrema ($DE_i$), black regions mark the unknown $DE_i$ value; (f) – pseudo-local $DE_i$ values, which are set as $V_i$. Pink and red rectangles in (a), (e) and (f) mark exemplary found regions of high (pink) and low (red) fringes period.

By virtue of a more robust calculation of the $V_i$ values, the $m_i$ is estimated in a much more predictable way. Thanks to that, based on the results of filtering many different single-frequency fringe patterns, we were able to set the parameter $\xi = 1.8$ and $\alpha = 30$ as default (these values enabled to obtain proper fringe pattern filtration for a satisfactorily wide range of frequencies). In practice, those parameters allow to obtain fringe pattern filtration that is close to optimal and therefore does not require user-driven adjustments.

The $V_i$ values in the FIF2 algorithm tend to be underestimated, as in general, implemented in FIF2 1D vector approach is oversensitive in finding the extrema locations in a 2D signal. This results in the production of a huge number of redundant BIMFs. The fpFIF2 algorithm does not struggle with this problem and due to novel, more reliable and 2D suitable method for $V_i$ calculation, produces smaller number of BIMFs. This simplifies the results and reduces the algorithm computation time without affecting the final filtration quality (what we further proved quantitatively in Section 3.1). Additionally, it is worth mentioning that the FIF2 algorithm by default assumes that the input image needs to be extended over its borders to reduce filtration errors on its edges. However, padding the fringe pattern image is not a trivial issue. The methods implemented in the FIF2 do not help to minimize the border errors (fringes become discontinuous, which introduces spurious higher frequency components and results in decomposition errors). By omitting this step in the fpFIF2 algorithm, we were still able to obtain similar (or even better, as we further show in Fig. 4) filtration accuracy at the image borders, without increasing the image domain, what allowed us to further decrease the fpFIF2 decomposition time.

*2.2. Additional BIMFs filtration*

Despite robust filter size calculation, the fpFIF2 algorithm may still exhibit mode mixing, mainly between noise and fringes information, exemplifying case is analyzed in Fig. 3. This may happen especially, when at some point of the fpFIF2 algorithm, in the $I_i$ image (Fig. 3(a)) there is only a small amount of noise left ($N_s$). In such a case, this noise may not significantly influence the $V_i$ values and the $m_i$ tends to be similar to noise-free cases. The BIMF$_n$ would then store signal frequencies that result from $m_i$ and higher, which means that $N_s$ would also be present in this mode, see Fig. 3(b).

To prevent such issues, we propose to produce additional BIMFs, to increase the BIMFs frequency sampling ("frequency resolution") and therefore increase the chance to decompose the $N_s$ into separate BIMFs. This extra filtration splits a certain BIMF into 2 separate BIMFs, see Fig. 3(c) and 3(d). In the former, the frequencies stored are higher than one resulting from the $m_i$ and in the latter, the frequencies stored correspond to the calculated $m_i$ value. To do not needlessly increase the number of produced BIMFs, we perform this additional filtration only when $m_i$ is at least twice as large as $m_{i-1}$, which is the case when the beneficial influence of additional filtering is the most probable. In practice, this usually translates into producing only between 1 to 4 additional BIMFs.

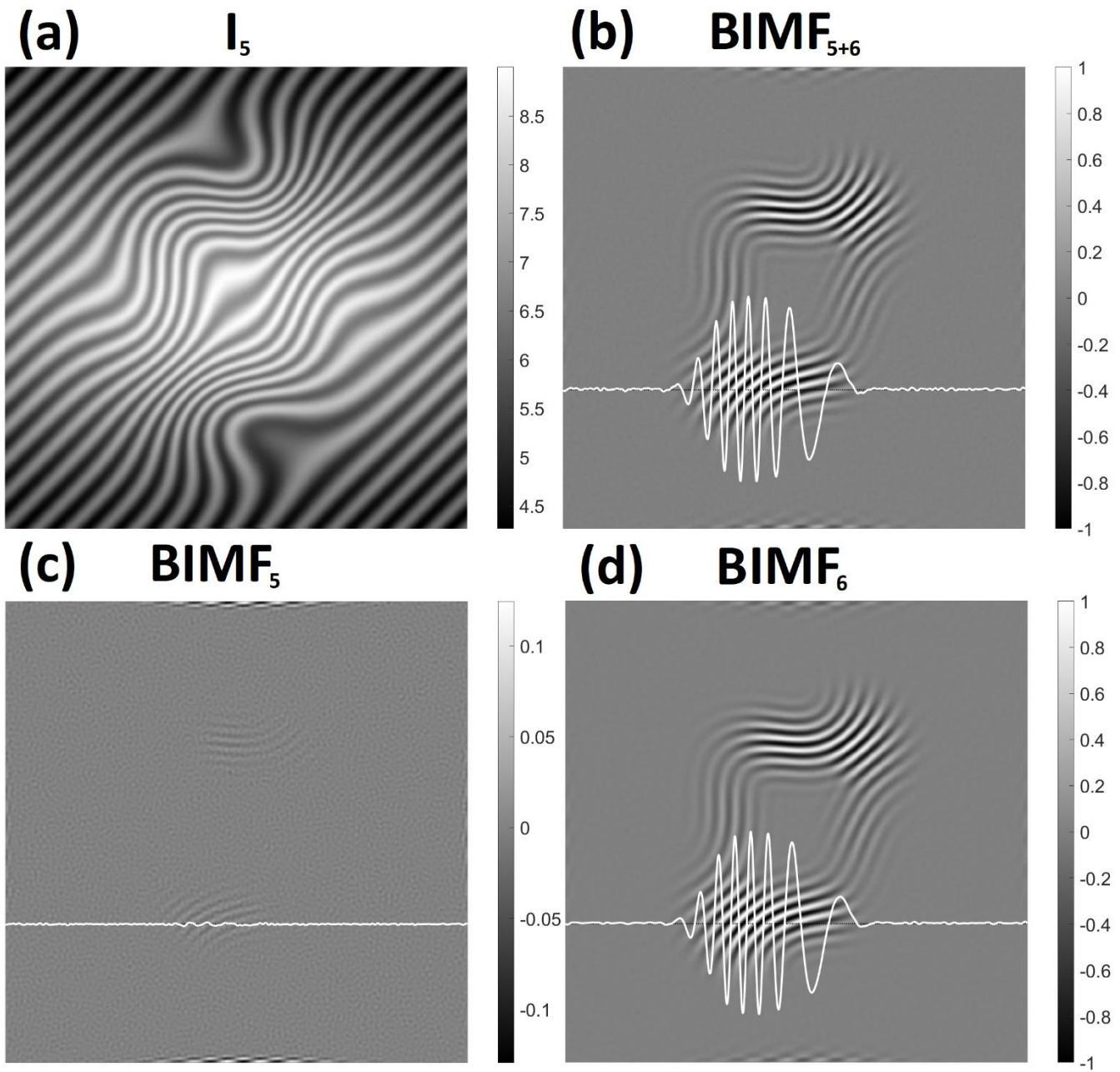

**Fig. 3.** Example of the beneficial influence of the additional filtration. (a) – Input image; (b) – one of the BIMFs that would be decomposed without additional BIMFs filtration (that corresponds to sum of BIMF$_5$ and BIMF$_6$ with additional filtration); (c) – auxiliary BIMF$_5$ that was produced by our additional filtration algorithm; (d) – BIMF$_6$, which thanks to additional filtration is free from the noise present in (c). White lines in (b-d) represent images cross-sections.

### 2.3. Additional stopping criteria of the algorithm's outer loop

FIF2 algorithm initially has only one outer loop stopping criterion. The algorithm stops when the number of extrema in the $I_i$ is smaller than 3. Since the number of extrema is computed in 1D (see Section 2.1), usually more extrema are found than are actually present in the 2D signal. This error produces an unnecessarily large number of modes. Improving the image extrema identification in the fpFIF2 (see Section 2.1) made this stopping criterion terminating the algorithm without producing a large number of low-frequency BIMFs. However, this criterion may not be sufficient in the case of an input image with a very complex background (having many local extrema). To address this problem, we added two novel stopping criteria (the second and the third stopping criterion in Fig. 1).

The second outer loop stopping criterion finishes the algorithm when the calculated filter size (Eq. (5)) is larger than the size of the image's larger dimension ($S_{xy}$). Thanks to that, the algorithm will stop when only low spatial frequencies are left in the $I_i$ (those frequency terms, with a high degree of probability, belong to the input image background components):

$$Stop\ if: m_i > S_{xy}. \tag{8}$$

The third outer loop stopping criterion was dedicated to stopping the algorithm when the input image contains as background component a constant value or it has no background component at all. This new stopping criterion terminates the fpFIF2 algorithm when the current image intensity is smaller than 1% of the input image intensity:

$$Stop\ if: \sum(I_i - \min(I_i)) < 0.01 \cdot \sum(I_1 - \min(I_1)), \tag{9}$$

where $\sum(x)$ represents the sum of all pixels of $x$ image.

## 3. Numerical experiments

In this section, we present the results of the numerical experiments. We compare the fpFIF2 algorithm to the previous 2D FIF implementation in terms of fringes filtration (Section 3.1) and to several reference algorithms with regards to fringes filtration and aiding HST algorithm in phase retrieval (Section 3.2).

### 3.1. Comparison of the fpFIF2 and FIF2 algorithms

To compare the fpFIF2 and FIF2 algorithms, we simulated a fringe pattern image (Fig. 4(a)) composed of background, noise, and fringes as in Eq. (10):

$$I(\vec{r}) = B(\vec{r}) + C \cdot \cos\left(\varphi(\vec{r}) + \vec{r} \cdot \frac{2\pi}{T}\right) + N(\vec{r}), \tag{10}$$

where $\vec{r}$ is a vector of image x-y coordinates (pix), $B(\vec{r})$ is the Gaussian distribution background (Fig. 4(d3)), $C$ is the image contrast level, $\varphi(\vec{r})$ is the object phase, T is the carrier fringes period and $N(\vec{r})$ is the normal distribution noise (Fig. 4(d1)). Then, we decomposed this image with fpFIF2 and FIF2 algorithms (we used default $\xi$ and $\alpha$ parameters). Next, by summation of relevant BIMFs we retrieved the noise, background and fringes information (Fig. 4(e1-e3) and 4(f1-f3)). Additionally, we compared the recovered fringes with the ground truth (Fig. 4(d2)) using a quantitative metric of structural similarity index measure (SSIM) [42], which returns the general similarity between two images on 0-1 scale. The maps of local SSIM indicators (calculated for each pixel) are shown in Fig. 4(b) and 4(c) and the total SSIM values were equal to: 0.995 for fpFIF2 and 0.990 for FIF2. According to both SSIM values and visual observations, fringes filtered with fpFIF2 algorithm are of much better quality. In the FIF2 decomposition, mode-mixing is more significant (noise information leaks into fringe component and fringe information into background mode). Thus, additional adjustment of $\xi$ and $\alpha$ is required. On the other hand, in the fpFIF2, only a negligibly small amount of fringe information is transferred to the noise component (Fig. 4(f1)), proving the near-ideal separation of the image components. This small mode-mixing effect should not have significant impact on phase retrieval from a fringe pattern, as show our further simulations (Section 3.2). Moreover, the FIF2 reconstruction exhibits lower decomposition quality in near-edge (image border) regions, see SSIM maps in Fig. 4(b) and 4(c). An additional advantage of the fpFIF2 algorithm is a much smaller number of returned BIMFs (14 for fpFIF2 and 37 for FIF2) and a much shorter decomposition time (this 500x500 pix image decomposition performed on a low-cost laptop lasted 0.8 s for fpFIF2 and 30 s for FIF2 algorithms).

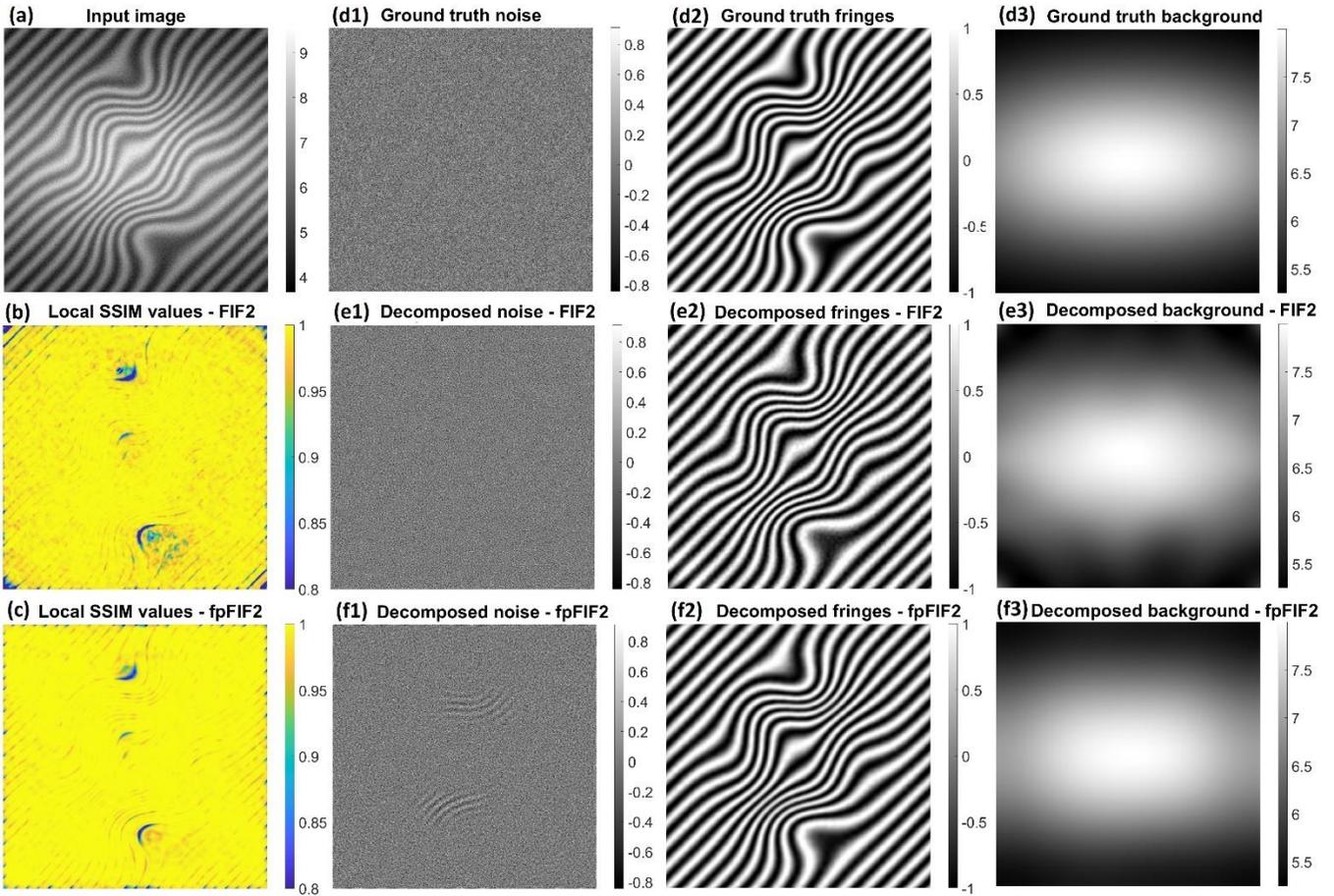

**Fig. 4.** Results of image filtering. (a) – input image composed of (d1) noise, (d2) fringes and (d3) background; (b), (c) – local SSIM values for (b) FIF2 and (c) fpFIF2 decompositions. Total SSIM was equal to 0.995 for fpFIF2 and 0.990 for FIF2; (e), (f) – decomposed noise, fringes and background for (e) FIF2 and (f) fpFIF2 algorithms.

To get more general results, we have simulated according to Eq. (10) a set of twenty-five 500x500 pix randomly generated fringe patterns (with random SNR varying from 9 to 20 dB; Gaussian background distribution with random sigma in x and y directions varying from 50 to 250; random PDR varying from 0.1 to 1; and fringe pattern carrier frequency with a random period in x and y direction, varying from 16 to 116 pix) and decomposed each of those images with both methods. For each decomposition, we have calculated: (1) decomposition time, (2) number of BIMFs and (3) SSIM similarity between decomposed fringes and ground truth for the 300x300 pix region in the image center (to do not take into account errors produced near image borders). Obtained results (Table 1) confirm that fpFIF2 achieves better fringe pattern filtration quality and returns a much smaller number of BIMFs in a much shorter time.

**Table 1**
Comparison of the results obtained by the FIF2 and fpFIF2 algorithms.

| Method | SSIM | | | Number of BIMFs | | | Decomposition time [s] | | |
|---|---|---|---|---|---|---|---|---|---|
| | min | average | max | min | average | max | min | average | max |
| FIF2 | 0.987 | 0.996 | 0.999 | 35 | 46.7 | 52 | 29.07 | 45.59 | 90.20 |
| fpFIF2 | 0.993 | 0.998 | 1 | 9 | 12.3 | 14 | 0.47 | 0.73 | 0.84 |

## 3.2. Comparison of the fpFIF2 with well-established fringe-decomposing algorithms

It is essential to test the new fringe preprocessing method versus pervasive reference methods. To evaluate the fpFIF2 algorithm numerically, we simulated the image composed of fringes, normal distribution random noise (with low SNR equal 7 dB) and Gaussian distribution background, according to Eq. (10). The fringe pattern is depicted in Fig. 5(a). Then, we extracted from it the fringe term with the use of: fpFIF2 (Fig. 5(c)) and three reference algorithms: EFEMD (Fig. 5(d)), EFEMD+BM3D (Fig. 5(e)), and uVID (Fig. 5(f)). Additionally, we compared the decomposed fringe terms with the ground truth simulated fringe cosine term (Fig. 5(b)), using SSIM measure. Maps of local SSIM values are shown in Fig. 5(c2)-5(f2), overall SSIM values are shown in captions. We did not include the FIF2 algorithm in this comparison as it can provide similar results to fpFIF2 but requires manual adjustment of $\xi$ and $\alpha$ parameters and consumes much more time for processing and selecting appropriate modes (as their number is significantly higher).

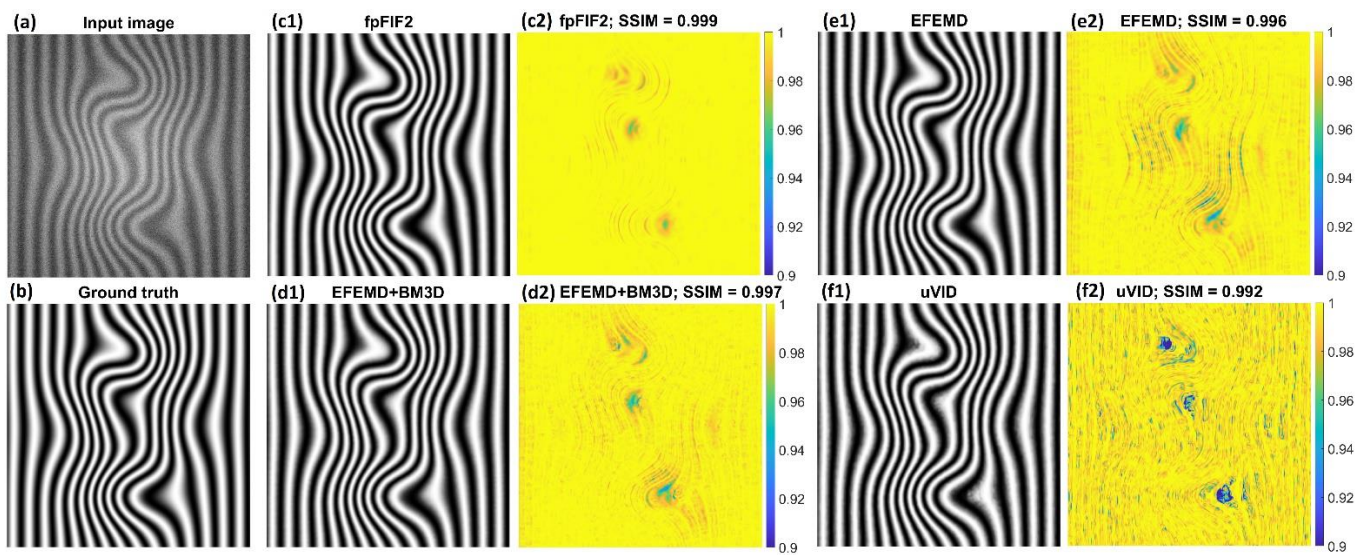

**Fig. 5.** Filtration of the exemplary fringe pattern image. (a) – simulated image; (b) – ground truth fringe pattern; (c1-f1) – fringe patterns decomposed by different methods. (c2-f2) – local SSIM values for tested methods. (c) – fpFIF2, (d) – EFEMD, (e) – EFEMD+BM3D, (f) – uVID.

Fig. **5** proves that the shape of the fringes decomposed by fpFIF2 is much better preserved than for reference algorithms (see SSIM maps in Fig. 5). The total SSIM value also quantitatively confirms this observation. Moreover, contrary to EFEMD and uVID algorithms, the fpFIF2 fringes seem not to be disturbed by any noise. Another advantage of the fpFIF2 algorithm is the short decomposition time, which is around three times shorter than for the second-fastest EFEMD algorithm (e.g., 2000x2000 pix image decomposition shown in Fig. 5(a) lasted 9 s for fpFIF2, 27 s for EFEMD, 48 s for EFEMD+BM3D and 2796 s for uVID algorithms). Moreover, because of decomposing images using convolutional filters, fpFIF2 can perform fast filtering of the series of many similar images (e.g., time-series images). In this case, there is a need to decompose only one of the images with the fpFIF2 and then use produced filters to decompose the rest. In such a way, for the image shown in Fig. 5(a), filtration using ready-made filters lasted only 0.31 s, which is a fantastic acceleration compared to all other well-established decomposition methods. Now, the only fpFIF2 limitation compared to other methods, is the relatively large number of returned BIMFs (e.g., for the image shown in Fig. 5(a), the fpFIF2 returned 15, EFEMD 8, EFEMD+BM3D 6 and uVID 3 BIMFs). Nonetheless, this disadvantage does not affect the quality of fringe decomposition and should not make it significantly more complicated to automatize than the EFEMD algorithm (e.g., using an automated selective reconstruction algorithm [20]). It may only be a little more complicated for the users as it forces them to decide which modes contain viable fringes basing on larger number of BIMFs than for other algorithms.

Proper fringe pattern filtration does not necessarily translate into high-quality HST phase reconstruction (e.g., a slight disturbance in filtered fringes shape may cause noticeable errors in the reconstruction [13, 15]). Because of that, it would be more meaningful in terms of quantitative phase imaging QPI modality to evaluate the decomposition methods based on the reconstructed phase quality and this approach will be used in this article.

Firstly, we investigated how the tested decomposition algorithms perform with reduced contrast of the fringe pattern images, see Fig. 6, as it is a ubiquitous experimental challenge and main factor limiting accuracy and applicability of QPI units. To do that, we simulated a set of fringe patterns with different contrast levels according to Eq. (10), with $C$ parameter changing from 20 to 100% and underlying phase distribution imitating a group of five HeLa cells, Fig. 6(b). An exemplary simulated image (for $C = 45\%$) is shown in Fig. 6(a). Then we decomposed all simulated images with the use of the fpFIF2, EFEMD, EFEMD+BM3D and uVID methods. After that, we supplied all filtered images to the HST algorithm to recover the simulated object phase (Fig. 6(c)-6(f)). We used the SSIM metric to compare the recovered phases with the phase (Fig. 6(b)) retrieved from the ideal fringe pattern image (simulated as in Eq. (10), with $B(\vec{r})$ and $N(\vec{r})$ parts equal to 0). In the end, we received the dependence between the simulated input image contrast and the SSIM quality of the reconstructed phase image for each of the decomposition methods, Fig. 6(g).

The results in Fig. 6 show that in a low-contrast regime (below 55%), the highest quality phase is obtained for the fpFIF2 algorithm. For higher contrasts, the EFEMD+BM3D method performance is on par with the fpFIF2 method. Nevertheless, the performance of these algorithms is still noticeably better than EFEMD and uVID methods and it is worth noting that fpFIF2 is a single stand-alone algorithm whereas EFEMD+BM3D constitutes a merger of two different algorithms. To evaluate the outcome of all algorithms, we set the SSIM value equal to 0.95 as a threshold value, above which the phase reconstruction is of good quality (black dotted line in Fig. 6(g)). According to this threshold, the lowest contrast value that allows for proper fpFIF2+HST phase reconstruction is equal to 30-35%, which is about 20-25% lower than for the second-best EFEMD+BM3D method. This numerical study indicates that fpFIF2 can algorithmically account for lowered coherence of light source, which physically limits the fringe pattern contrast and lowers the quality of measured images.

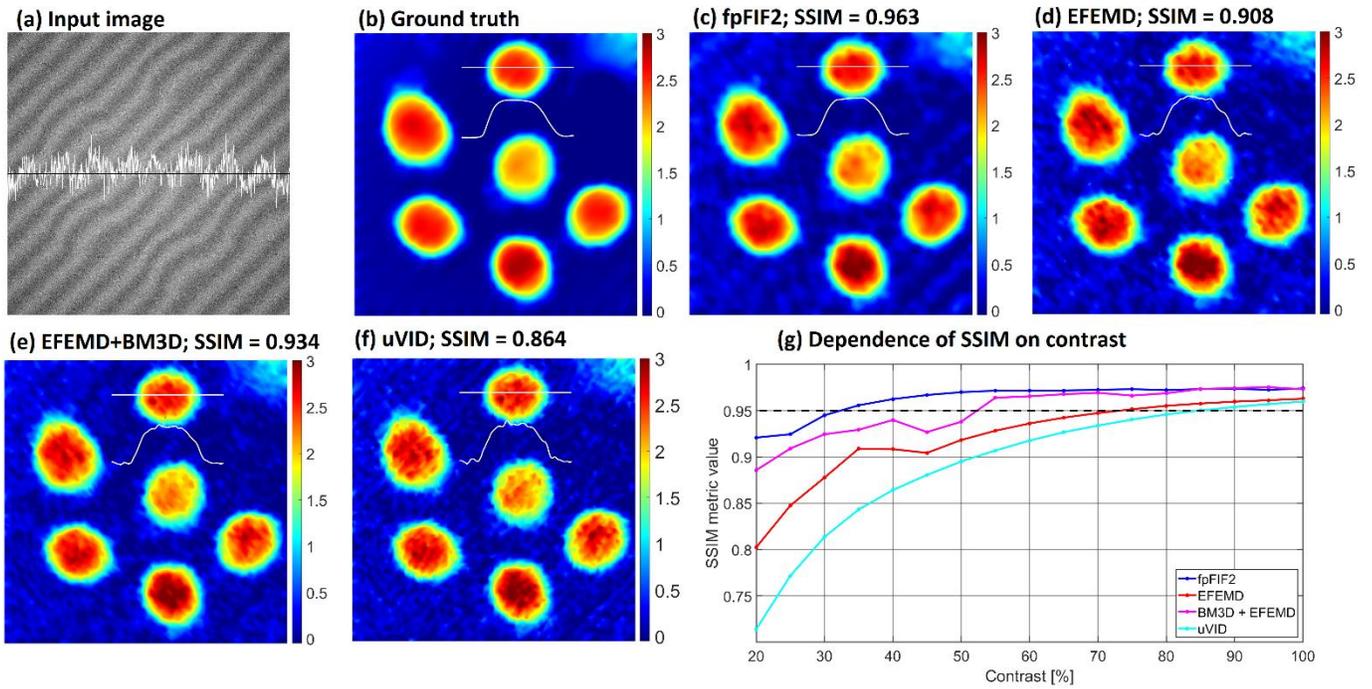

**Fig. 6.** Numerical experiments – reconstruction quality as a function of image contrast level. (a) – simulated fringe pattern with contrast level equal to 45%. (b) – object phase reconstructed from ground truth fringe pattern. (c)-(f) – HST phase reconstruction of the fringe pattern filtered with: (c) fpFIF2, (d) EFEMD, (e) EFEMD+BM3D and (f) uVID methods. (g) – SSIM metric values of the phase reconstruction for different input image contrast values.

Fringe pattern decomposition performance may enormously vary depending on fringe pattern parameters. Hence the above results may slightly differ for a range of simulation parameters (i.e., different fringe pattern carrier frequency and spatially varying fringe period/orientation). To consider that, we performed two additional numerical experiments with analogous processing patches as before to test the performance of the fpFIF2 algorithm with low SNR = 7 dB data for various input image conditions, see Fig. 7.

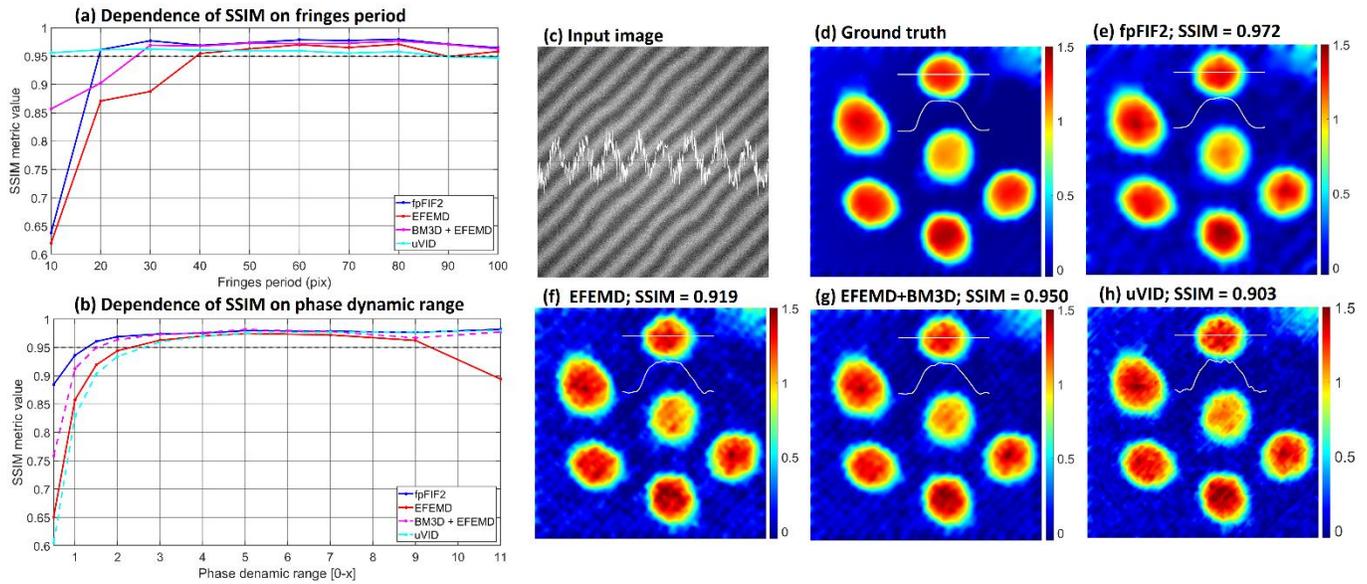

**Fig. 7.** Numerical experiments: SSIM metric values for different carrier fringes period (a) and different phase dynamic range (b). (c) – input image with low PDR = 1.5; (d) – ground truth phase; (e)-(h) – phase reconstructed with (e) fpFIF2, (f) EFEMD, (g) EFEMD+BM3D and (h) uVID methods.

In the first of those experiments, investigated algorithms filtered a set of fringe patterns simulated with different periods of the carrier fringes (T parameter in Eq. (10)). As a result, we received the SSIM phase quality values as a function of the signal carrier period, see Fig. 7(a). These results show that, apart from the smallest tested signal period, the fpFIF2+HST reconstructed phase quality was similar or better than for reference methods.

In the second experiment, we tested the algorithm performance for different phase dynamic ranges. The higher is the PDR, the curvier the fringes are in local object-present areas, and therefore, the fringe pattern is composed of the higher number of spatial frequency components. In theory, high PDR fringes should be more challenging to filter (the algorithms need to determine which image frequencies are a part of a signal), whereas low PDR fringes require more precise filtering in order to detect very small fringe curvature (in other words, to preserve their small local deviations from straightness). The results of this experiment, Fig. 7(b), show that for the phase dynamic range 3-11, the fpFIF2 performance is similar or slightly better than other algorithms. For lower PDRs (exemplary phase reconstruction for PDR = 1.5 is shown in Fig. 7(c)-7(h)), the fpFIF2 SSIM metric values are noticeably higher than other algorithms, which shows that fpFIF2 is more sensitive to small object phase changes.

## 4. Experimental results

To confirm the results of these promising simulations, we have performed two real data experiments. In the first one we used a classical QPI Mach-Zehnder interferometer (light source – laser diode emitting at 635 nm, objective – 20X with a numerical aperture of 0.75, camera – 4000x3000 pix with pixel pitch 1.85 µm) to collect the interferograms of a biological sample – a specimen containing a group of human adipose-derived stem cells (hADSCs) isolated from fat tissue, see Fig. 8. Cells were cultured in a dedicated medium (RoosterBio). Cells that reached appropriate density on glass coverslips were fixed in 4% paraformaldehyde and mounted on the microscope slides in fluorescent mounting medium (DAKO). Firstly, we collected a series of 6 high-contrast (82%), Fig. 8(b), randomly phase-shifted interferograms and one low-contrast interferogram (contrast was equal to 33% – Fig. 8(c)), where low contrast was introduced by placing a polarizer in the interferometer reference beam. After that, we used the PCA algorithm [7] on the series of high-contrast phase-shifted interferograms to obtain the cells' reference phase, Fig. 8(a), and performed a single-frame HST reconstruction on filtered high- and low-contrast interferograms, Fig. 8(d)-8(g). For each of the obtained phase images, we have calculated the SSIM metric values (see Fig. 8), comparing a small region (areas inside red boxes in Fig. 8) of the reconstructed phases with the reference phase.

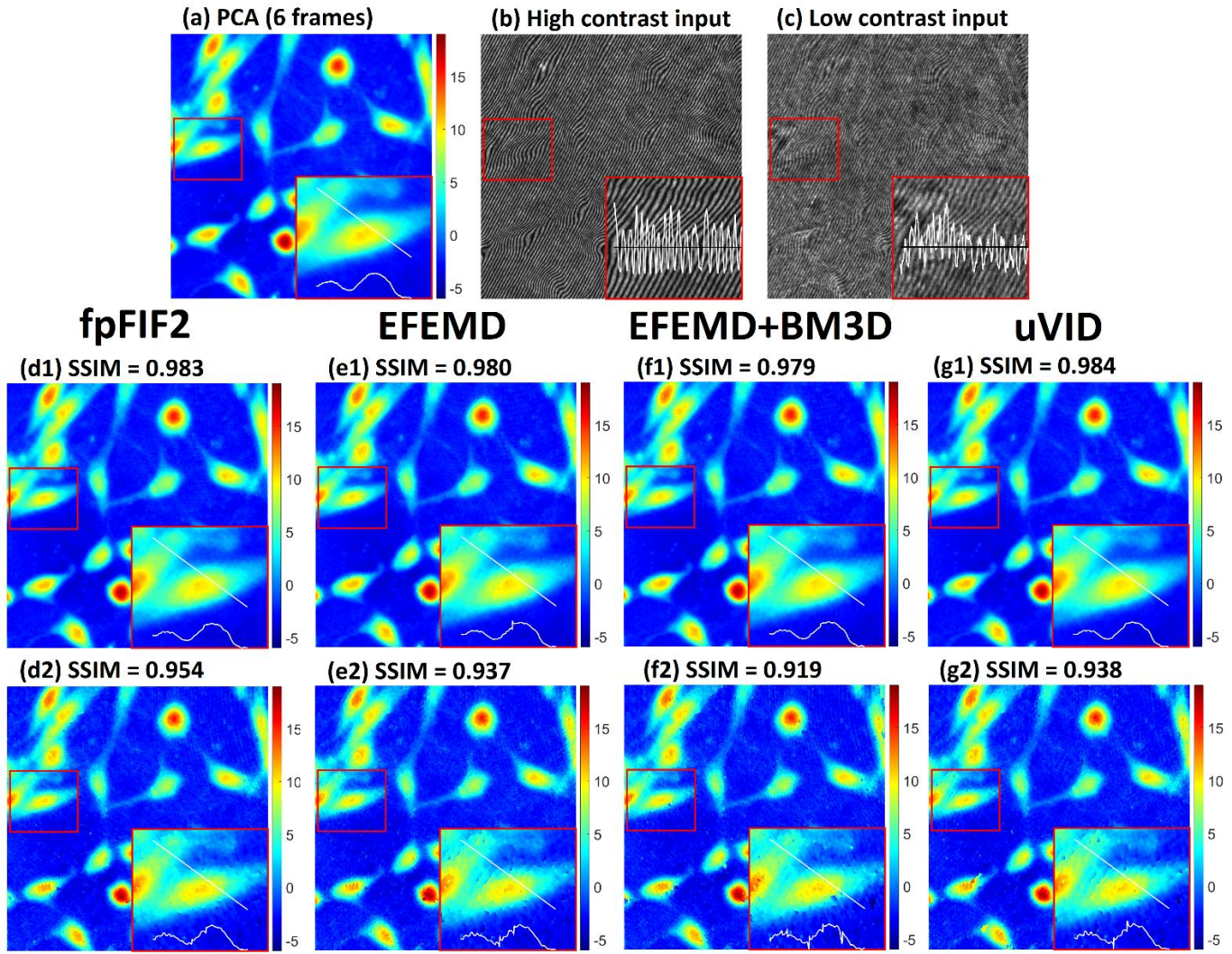

**Fig. 8.** Experimental validation – cells phase reconstruction. (a) – reference phase reconstructed by the PCA method. (b) – high contrast input fringe pattern. (c) – low contrast input fringe pattern. (d)-(g) HST reconstruction of the filtered: (d1)-(g1) high-contrast and (d2)-(g2) low-contrast fringes. (d) – fpFIF2 filtration, (e) – EFEMD filtration, (f) – EFEMD+BM3D filtration, (g) – uVID filtration.

For high-contrast input images, each reconstructed phase image can be visually assessed to have similarly good quality. However, the SSIM values for fpFIF2 and uVID reconstructions are larger by 0.003 than for other algorithms. For low-contrast images, the fpFIF2 reconstruction is observed to outperform the other methods. The fpFIF2+HST reconstructed phase has the smallest number of local phase discontinuities and the cells profiles, represented by the white lines in Fig. 8, are the smoothest. Moreover, for the fpFIF2, the SSIM is larger by 0.016 than for the second-best uVID algorithm.

In the second experiment, Fig. 9, we measured a technical sample – a phase test target composed of rectangles of 3 μm height, xy dimensions varying from 2.5x2.5 to 5x5 μm$^2$ and refractive index equal to 1.552. We used similar interferometric setup as in the first experiment, but with a 100X objective with 0.9 numerical aperture.

Firstly, we deployed it to record a single, high contrast fringe pattern image of high fringes density, which was used to obtain object reference phase by the Fourier transform method, Fig. 9(a). Then, we collected 2 fringe patterns of lower fringe density: one with higher contrast (63%), Fig. 9(b), and one with lower contrast (41%), Fig. 9(c). Afterwards, we filtered those fringes with methods mentioned in this article and used the HST to demodulate the object phase, Fig. 9(d)-9(g). For each of the obtained phase images, we have calculated the SSIM metric values (see Fig. 9), comparing a small region (areas inside red boxes in Fig. 9) of the reconstructed phases with the reference phase.

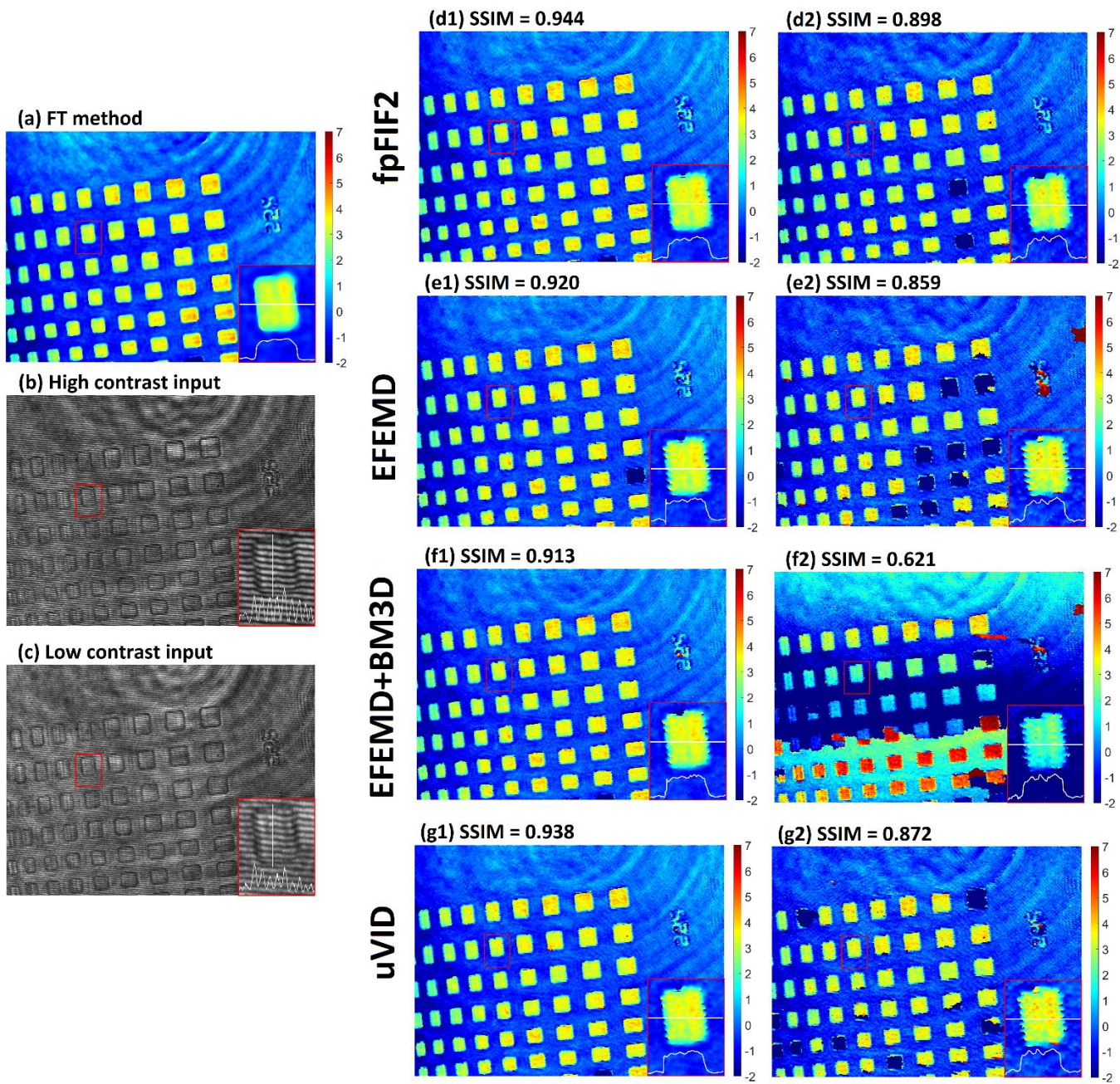

**Fig. 9.** Experimental validation – phase test target reconstruction. (a) – reference phase reconstructed by the FT method. (b) – high contrast input fringe pattern. (c) – low contrast input fringe pattern. (d)-(g) HST reconstruction of the filtered: (d1)-(g1) high-contrast and (d2)-(g2) low-contrast fringes. (d) – fpFIF2 filtration, (e) – EFEMD

filtration, (f) – EFEMD+BM3D filtration, (g) – uVID filtration. Color scales in (a), (d)-(f) correspond to retrieved phase values, which can be converted to sample height by multiplying it by 0.7 factor.

As it can be observed, the phase of the technical sample is more challenging to retrieve than the biological sample one, mainly due to sharp changes in the object phase, which are very sensitive to single-shot phase demodulation and spatial unwrapping errors. Comparing the reconstructions basing on unwrapping errors, for high contrast images, fpFIF2, EFEMD and uVID algorithms obtained no errors, whereas the EFEMD provided erroneous reconstruction for one rectangle (dark blue rectangle in Fig. 9(e1)). For lower contrast images, the fpFIF2 wrongly reconstructed only 2 rectangles, whereas EFEMD and uVID had not coped well with 10 and 8 rectangles, respectively. On the other extreme is EFEMD+BM3D low contrast reconstruction, where phase unwrapping errors damaged the whole reconstructed field of view. Apart from the number of unwrapping errors, other factors also work for the favor of fpFIF2 algorithm. The edges of the reconstructed rectangles seem to be the smoothest for fpFIF2 reconstructions, whereas others are spoiled by fringe-like errors, what is especially seen for low contrast inputs. Moreover, the SSIM values calculated for a single, correctly unwrapped rectangle are also highest for fpFIF2 for both high and low contrast input fringe pattern images.

## 5. Summary

To sum up, we have proposed the fpFIF2 – a novel algorithm for fringe pattern filtration aiding the HST phase recovery technique, crucial in high phase space-time-bandwidth QPI product. We have shown that our algorithm is better tailored for fringe pattern processing than the previous and general signal processing 2D FIF implementation. Moreover, our synthetic and experimental data analysis provided quantitative evidence that, independently from the fringe pattern parameters, the fpFIF2 influence onto the HST algorithm is at least as beneficial as other tested decomposition algorithms. Most importantly, in low SNR and PDR regimes, HST performance with the fpFIF2 prefiltering is noticeably better, both qualitatively (visually) and quantitatively (SSIM), than with the other algorithms. Additionally, the fpFIF2 performs faster than all compared methods and it may be easily adapted to fast filtering of large time-series datasets (with predefined filters). All highlighted features indicate that fpFIF2 may be a valuable tool for optical fringe pattern filtering and may become an exciting alternative to BEMD/VID-based algorithms, especially in a very challenging fringe pattern low SNR regime.

**Acknowledgments.** We would like to thank Maria Cywińska for sharing the uVID algorithm codes and Piotr Arcab and Emilia Wdowiak for lending the phase test target. The authors declare no conflicts of interest.

**Funding.** This work was financed by National Science Center, Poland (2020/37/B/ST7/03629), Foundation for Polish Science, Poland (TEAM-NET POIR.04.04.00-00-16ED/18) and by NanoTech4ALS (ref. ENMed/0008/2015, 13/EuroNanoMed/2016). Antonio Cicone is a member of the Italian National Group for Scientific Computation (GNCS-INDAM), and he thanks the Italian Space Agency, for the financial support under the contract ASI "LIMADOU scienza +" n. 2020-31-HH.0, and the ISSI-BJ project "the electromagnetic data validation and scientific application research based on CSES satellite" and Dragon 5 cooperation 2020-2024 (ID. 59236).